\documentclass[draftcls, onecolumn]{IEEEtran}
\ifCLASSINFOpdf
\else
\fi
\usepackage[cmex10]{amsmath}
\usepackage{array}

\usepackage{mdwmath}
\usepackage{mdwtab}
\usepackage{fixltx2e}
\usepackage{graphicx}
\usepackage{amssymb}
\usepackage{amsbsy} 
\usepackage{float}
\usepackage{balance}
\usepackage{subfigure}
\usepackage{url}
\newfloat{longequation}{b}{ext}

\newtheorem{thm}{Theorem}[section]

\newtheorem{lem}{Lemma}[section]
\newtheorem{pro}{Property}[section]
\newtheorem{df}{Definition}[section]
\newtheorem{rem}{Remark}[section]

\begin{document}

%
\title{Free Probability based Capacity Calculation \\of Multiantenna Gaussian Fading Channels \\with Cochannel Interference}
%
%
%

\author{Symeon~Chatzinotas,
        Bj\"{o}rn~Ottersten
\thanks{S. Chatzinotas and B. Ottersten are with the Interdisciplinary Centre for Security, Reliability and Trust, University of Luxembourg  (http://www.securityandtrust.lu)
 e-mail: \textbraceleft Symeon.Chatzinotas, Bjorn.Ottersten\textbraceright@uni.lu.}
}

\maketitle

\begin{abstract}
During the last decade, it has been well understood that communication over multiple antennas can increase linearly the multiplexing capacity gain and provide large spectral efficiency improvements. However, the majority of studies in this area were carried out ignoring cochannel interference. Only a small number of investigations have considered cochannel interference, but even therein simple channel models were employed, assuming identically distributed fading coefficients. In this paper, a generic model for a multi-antenna channel is presented incorporating four impairments, namely additive white Gaussian noise, flat fading, path loss and cochannel interference. Both point-to-point and multiple-access MIMO channels are considered, including the case of cooperating Base Station clusters. The asymptotic capacity limit of this channel is calculated based on an asymptotic free probability approach which exploits the additive and multiplicative free convolution in the $R$- and $S$-transform domain respectively, as well as properties of the $\eta$ and Stieltjes transform. Numerical results are utilized to verify the accuracy of the derived closed-form expressions and evaluate the effect of the cochannel interference.
\end{abstract}

\begin{IEEEkeywords}
Information theory, Information Rates, Multiuser channels, MIMO systems, Cochannel Interference, Land mobile radio cellular systems, Eigenvalues and eigenfunctions.
\end{IEEEkeywords}

%
\IEEEpeerreviewmaketitle

\section{Introduction}
\IEEEPARstart{I}{n} 
many cases, wireless communication systems have to operate in the interference-limited regime, where the cochannel interference is much more pronounced that the receiver noise. This applies to all modern cellular systems, as well as in multi-spot beam satellites, where frequency reuse is employed over spatially separated geographical areas. The capacity of those systems is interference-limited, since by increasing the transmit power both received signal and cochannel interference increase and eventually the Signal over Interference and Noise Ratio (SINR) saturates. Although a number of previous studies in the literature \cite{Blum2003,Lozano2002,Moustakas2003,Choi2008,Somekh2007} have looked into the effect of cochannel interference, the majority of the employed channel models were focused on specific cases, failing to capture the wide range of affected systems.

In this context, this paper generalizes and studies the effect of cochannel interference on multiantenna Gaussian fading channels. More specifically, the main contributions herein are:
\begin{enumerate}
        \item The introduction of a generic multidimensional channel model which encompasses four channel impairments, namely additive white Gaussian noise (AWGN), flat fading, path loss and cochannel interference. The proposed model can be employed as a generalization of point-to-point MIMO channels, as well as uplink and downlink channels of MIMO cellular systems, cooperating Base Station clusters and multi-spot beam satellite systems. 
        \item The analytical calculation of the asymptotic eigenvalue probability distribution function (a.e.p.d.f.) based on a free probability approach which exploits the additive and multiplicative free convolution \cite{Tulino04} in the $R$- and $S$-transform domain respectively, as well as properties of the $\eta$ and Stieltjes transform. 
        \item The derivation of closed-form methods which calculate the system capacity based on the number of dimensions and the transmit power of useful signals and cochannel interference. 
        \item Numerical results which verify the validity of the free probability derivations and provide insights into the capacity performance of cochannel-interfered systems.
\end{enumerate}

The remainder of this paper is structured as follows: Section \ref{sec: related work} introduces the generic channel model and provides a detailed review of cochannel interference scenarios. Section \ref{sec: eigenvalue distribution analysis} describes the free probability derivations and the capacity results, while cumbersome mathematical derivations are postponed to the appendix. Section \ref{sec: numerical results} verifies the accuracy of the analysis by comparing with Monte Carlo simulations and evaluates the effect of the cochannel interference in the context of cooperating BS clusters. Section \ref{sec: conclusion} concludes the paper.

\subsection{Notation}
Throughout the formulations of this paper, \(\mathbb{E}[\cdot]\) denotes the expectation,  \(\left(\cdot\right)^H\) denotes the conjugate transpose matrix, \(\left(\cdot\right)^T\) denotes the transpose matrix, \(\odot\) denotes the Hadammard product and \(\otimes\) denotes the Kronecker product. The Frobenius norm of a matrix or vector is denoted by \(\left\Vert \cdot\right\Vert\), the absolute value of a scalar is denoted by \(\left\vert \cdot\right\vert\) and the delta function is denoted by 
$\delta(\cdot)$. $(\cdot)^+$ is equivalent to $\max(0,\cdot)$ and \(1\left\{ \cdot \right\}\) is the indicator function.

\section{Generic Channel Model \& Related Work}
\label{sec: related work}

\subsection{Generic Channel Model}
The generic channel model which combines additive white Gaussian noise, flat fading, path loss and cochannel interference can be expressed as follows:
\begin{equation}
\mathbf{y}=\mathbf{H}\mathbf{x}+\mathbf{H}_\mathrm{I}\mathbf{x}_\mathrm{I}+\mathbf{z},
\label{eq: generic channel model}
\end{equation}
where $\mathbf{y}$ denotes the $K\times 1$ received symbol vector and the $K\times 1$ vector $\mathbf{z}$ denotes AWGN with $\mathbb{E}[\mathbf{z}]=\mathbf{0}$ and $\mathbb{E}[\mathbf{zz}^H]=\mathbf{I}$\footnote{The variance of AWGN has been normalized to $1$ to simplify notations}. The $M\times 1$ vector $\mathbf{x}$ denotes the  transmitted symbol vector with Signal to Noise Ratio (SNR) \(\mu\) ($\mathbb{E}[\mathbf{xx}^H]=\mu\mathbf{I}$), while the $N\times 1$ vector $\mathbf{x}_\mathrm{I}$ denotes cochannel interference with Interference to Noise Ratio (INR) \(\nu\) ($\mathbb{E}[\mathbf{x}_\mathrm{I}\mathbf{x}_\mathrm{I}^H]=\nu\mathbf{I}$). It should be noted that $\mu$ represents the transmitted desired signal over noise power ratio (TSNR), while $\nu$ represents the transmitted interference over noise power ratio (TINR). The $K\times M$ channel matrix $\mathbf{H}=\mathbf{\Sigma} \odot \mathbf{G}$ comprises of the Hadammard product of a Gaussian matrix $vec(\mathbf{G})\sim\mathcal{CN}(\mathbf{0},\mathbf{I})$ including the flat fading coefficients of the communication system and a variance profile matrix $\mathbf{\Sigma}$ including the path loss coefficients of the communication system. Similarly, the $K\times N$ channel matrix $\mathbf{H}_\mathrm{I}=\mathbf{\Sigma}_\mathrm{I} \odot \mathbf{G}_\mathrm{I}$ comprises of the Hadammard product of a Gaussian matrix $vec(\mathbf{G}_\mathrm{I})\sim\mathcal{CN}(\mathbf{0},\mathbf{I})$ including the flat fading coefficients of the cochannel interference and a variance profile matrix $\mathbf{\Sigma}_\mathrm{I}$ including the path loss coefficients of the cochannel interference. The exact structure of the variance profile matrices depends on the considered wireless scenario and it is discussed in detail in the following section\footnote{In the previous notation, it is assumed that TSNR is identical for all desired dimensions and the TINR for all interfering dimensions. If this is not the case, variations in the transmit power across the multiple dimensions can be incorporated in the variance profile matrices $\mathbf{\Sigma}$ and $\mathbf{\Sigma}_\mathrm{I}$ respectively.}. It is also assumed that:

\begin{itemize}
\item 
$\mathbf x$ and $\mathbf x_\mathrm I$ are Gaussian inputs 
\item Cochannel interference is treated as noise
\item 
Channel State Information (CSI) is available at the receiver but not at the transmitters. Therefore, no input optimization takes place in order to avoid cochannel interference.
\end{itemize}  

In the proposed model, the combined effect of receiver noise plus cochannel interference can be represented by a combined vector $\mathbf{z}_\mathrm{I}=\mathbf{H}_\mathrm{I}\mathbf{x}_\mathrm{I}+\mathbf{z}$ which is characterized as colored since its covariance $\mathbb{E}[\mathbf{z}_\mathrm{I}\mathbf{z}_\mathrm{I}^H]=\mathbf{I}+\nu\mathbf{H}_\mathrm{I}\mathbf{H}^H_\mathrm{I}$ is no longer proportional to the identity matrix. The capacity of the generic model normalized by the number of receive dimensions is given by an expression of the following form\footnote{It should be noted that eq. \eqref{eq: generic capacity} can be also written as $\mathrm{C}=\frac{1}{K}\mathbb{E}[\log\det(\mathbf{I}+\mu\mathbf{HH}^H+\nu\mathbf{H}_\mathrm{I}\mathbf{H}_\mathrm{I}^H)]-\frac{1}{K}\mathbb{E}[\log\det(\mathbf{I}+\nu\mathbf{H}_\mathrm{I}\mathbf{H}_\mathrm{I}^H)]$ as in \cite{Chatzinotas_arxiv}. However, this paper focuses on eq. \eqref{eq: generic capacity} in order to distinguish and exploit the structure of matrices $\mathbf{\Sigma}$ and $\mathbf{\Sigma}_\mathrm{I}$.} \cite{Foschini1998,Blum2003,Lozano2002}:
\begin{align}
\mathrm{C}&=\frac{1}{K}\mathbb{E}\left[\log\det\left(\mathbf{I}+\mu\mathbf{HH}^H\left(\mathbf{I}+\nu\mathbf{H}_\mathrm{I}\mathbf{H}_\mathrm{I}^H\right)^{-1}\right)\right].
\label{eq: generic capacity}
\end{align}
\begin{rem}
At this point, it should be noted that the variance profile matrices should satisfy a number of conditions for the analytical results of section \ref{sec: eigenvalue distribution analysis} to be valid:
\begin{itemize}
\item 
$\mathbf{\Sigma}$ and $\mathbf{\Sigma}_\mathrm{I}$ are assumed to have uniformly bounded entries with growing dimensions and satisfy Lindeberg's condition.
\item
$\mathbf{\Sigma}$ and $\mathbf{\Sigma}_\mathrm{I}$ are assumed to be asymptotically row-regular.
\end{itemize}
\end{rem}
\begin{df}
An $N\times K$ matrix $\mathbf{X}$ is asymptotically row-regular \cite{Tulino04} if
\begin{equation}
\lim_{K\rightarrow \infty}\frac{1}{K}\sum_{j=1}^K 1\left\{ X_{i,j}\leq \alpha \right\}
\end{equation}
is independent of \(i\) for all \(\alpha\in \mathbb{R}\), as the aspect ratio \(\frac{K}{N}\) converges to a constant.
\end{df}
To simplify the notations during the mathematical analysis, the following auxiliary variables are defined:
\begin{align}
\mathbf{K}&=\mu \mathbf{HH}^H\left(\mathbf{I}+\nu\mathbf{H}_\mathrm{I}\mathbf{H}_\mathrm{I}^H\right)^{-1}\nonumber\\
\mathbf{M}&=\left(\mathbf{I}+\nu\mathbf{H}_\mathrm{I}\mathbf{H}_\mathrm{I}^H\right)^{-1}\nonumber\\
\mathbf{N}&=\mu\mathbf{HH}^H\nonumber\\
\mathbf{\tilde{N}}&=\nu\mathbf{H}_\mathrm{I}\mathbf{H}_\mathrm{I}^H\nonumber\\
\beta&=\frac{M}{K}\nonumber\\
\gamma&=\frac{N}{K}\nonumber\\
q&=\frac{\left\Vert\mathbf{\Sigma}\right\Vert^2}{MK}\nonumber\\
p&=\frac{\left\Vert\mathbf{\Sigma}_\mathrm{I}\right\Vert^2}{NK}\nonumber
\end{align}
where $\beta,\gamma$ are the ratios of horizontal to vertical dimensions of matrix $\mathbf{H},\mathbf{H}_\mathrm{I}$ respectively and $q,p$ are the squared Frobenius norms of matrices $\mathbf{\Sigma},\mathbf{\Sigma}_\mathrm{I}$ respectively normalized by the matrix size. 

\subsection{Wireless Scenarios with Cochannel Interference}
This generic model can encompass a wide range of point-to-point and multiple-access MIMO channels by altering the structure of the variance profile matrices and the meaning of the channel matrix dimensions, as described in the following paragraphs.

\subsubsection{Single MIMO Link with Equidistant Cochannel Interference}
\label{subsubsec: scenario 1}
In this scenario, a single MIMO link is considered whereas the receiver is impaired by a) a single cochannel MIMO interferer or b) a number of equidistant interferers. An example for case (a) would be two point-to-point MIMO links operating in close proximity, while for case (b) a cell-edge multiple-antenna terminal at the downlink channel receiving interference from adjacent equidistant Base Stations (BSs). In the described scenarios, the path loss coefficients can be considered identical and therefore the variance profile matrices are matrices of ones: $\mathbf{\Sigma}=\mathbf{\Sigma}_\mathrm{I}=\mathbb{I}$. In addition, $K$ and $M$  equal the number of receive antennas and the number or transmit antennas at the desired terminal respectively. $N$ equals the number of transmit antennas at the interfering terminal in case (a) or the number of interfering BSs in case (b).
Authors in \cite{Moustakas2003} investigated the correlated MIMO capacity with correlated cochannel interference, as well as the optimum signaling for fading channels. Using complex integrals and Grassman variables, the mutual information moments have been derived and a method for optimizing the input signal covariance based on the correlation matrices was presented. It has been shown that in many cases, the input optimization yields capacities close to the closed-loop capacity, where instantaneous channel state information is available at the receiver.

\subsubsection{Single MIMO Link with Distributed Cochannel Interference}
\label{subsubsec: scenario 2}
In this scenario, a MIMO link operating within a traditional cellular system is assumed. In the uplink, each BS receives cochannel interfering signals from the User Terminals (UTs) of adjacent cells. Similarly, in the downlink each UT receives cochannel interference from adjacent BSs. The main differentiation between uplink and downlink is that in the former case interference originates from a large number of low-power randomly-distributed sources (UTs), whereas in the latter case interference originated from a small number of high-power regularly-distributed sources (BSs). In both uplink and downlink, the variance profile matrix $\mathbf{\Sigma}$ is a matrix of ones $\mathbf{\Sigma}=\mathbb{I}$, while the structure of matrix $\mathbf{\Sigma}_\mathrm{I}=\boldsymbol{\sigma}_\mathrm{I}^T\otimes \mathbb{I}$ is dictated by the spatial distribution of the interferers, where $\boldsymbol{\sigma}_\mathrm{I}$ is a vector including the path loss components of the interferers. Regarding the uplink channel matrix dimensions, $K$ and $M$ refer to the number of receive antennas at the BS and the number of transmit antennas at the UT respectively, while $N$ refers to the number of interfering UTs multiplied by the number of transmit antennas per UT. 
Authors in \cite{Lozano2002} have investigated this scenario and have derived an asymptotic closed-form expression as a function of $K,M,N,p,q$ for AWGN and fading channels. According to this study the spatial noise coloring due to cochannel interference can be exploited by the multiple-antenna receiver in order to achieve higher channel capacities. Subsequently, the author in \cite{Blum2003} has studied the optimum signaling in MIMO channels with cochannel interference.  In this context, it has been found that the optimal signaling converges to the interference-free signaling only when interference is sufficiently weak of sufficiently strong. 

\subsubsection{MIMO Cellular System with Wideband Transmission Scheme}
\label{subsubsec: scenario 3}
This scenario considers a single cell which operates on wideband transmission scheme (i.e. superposition coding) and receives interference from adjacent cells. The wideband transmission scheme implies that all UTs of the cell of interest transmit over the same channel dimensions using superposition coding \cite{Wyner1994,Somekh2000}. Assuming BS and UTs equipped with multiple antennas, the uplink can be represented by a MIMO multiple-access channel. In this case, the structure of matrix $\mathbf{\Sigma}=\boldsymbol{\sigma}^T\otimes \mathbb{I}$ is dictated by the spatial distribution of the intra-cell UTs, while $\mathbf{\Sigma}_\mathrm{I}=\boldsymbol{\sigma}_\mathrm{I}^T\otimes \mathbb{I}$ is dictated by the inter-cell interferers, where $\boldsymbol{\sigma}$ is a vector including the path loss components of the desired UTs. Regarding the channel matrix dimensions, $K$ refers to the number of receive antennas at the BS, $M$  refers and the number of UTs times the number of transmit antennas at the UT and $N$ refers to the number of interfering UTs times the number of transmit antennas per UT. 
This scenario has been studied in \cite{Dai2004}, where various multiuser MIMO processing techniques are considered in the presence of cochannel interference. Furthermore, it is shown therein that linear MMSE (Minimum Mean Square Error) filtering yields eq. \eqref{eq: generic capacity}, assuming full CSI at the receiver. 

\subsubsection{Cooperating BS Cluster}
\label{subsubsec: scenario 4}
In this scenario, a cluster of cooperating BSs (or distributed antennas) is considered, receiving interference from similar adjacent clusters. Many variations of this scenario can be described by the generic channel model. For example, the BSs and UTs may be equipped with single or multiple antennas. In addition, orthogonal (single transmitting UT per cluster) or wideband (multiple transmitting UTs per cluster) transmission scheme can be considered. In all aforementioned cases the variance profile matrices have a special structure defined by the spatial distribution of BS clusters and intra-cell/inter-cell UTs. The defining characteristic of all the aforementioned cases is that matrices $\mathbf{\Sigma}$ and $\mathbf{\Sigma}_\mathrm{I}$ can no longer be expressed in terms of the vector $\boldsymbol{\sigma}$ or $\boldsymbol{\sigma}_\mathrm{I}$. The reader is referred to \cite{Chatzinotas_Chapter1} for a detailed review of variance profile matrices for BS cooperation systems. Cooperating BS clusters were also considered in \cite{Choi2008}, using inter-cell scheduling to avoid cochannel interference. Furthermore, the authors in \cite{Somekh2007} consider BS clusters with frequency reuse 1 on a circular Wyner array, although all intra-cell UTs are assumed to have equal path loss coefficients in order to provide mathematical tractability. 
\begin{rem}
It should be noted that the generic model and capacity derivation in this paper generalize all the aforementioned scenarios by making no simplifying assumptions except the asymptotic row-regularity of the variance profile matrices. In terms of wireless scenarios, asymptotic row-regularity means that each receiving element collects in total the same power from all transmitters.
This is always true for multiple collocated omnidirectional receive antennas, such as scenarios \ref{subsubsec: scenario 1}, \ref{subsubsec: scenario 2}. However, for spatially distributed receive antennas such as in scenarios \ref{subsubsec: scenario 3}, \ref{subsubsec: scenario 4} asymptotic row-regularity translates into symmetric cells in terms of spatial user distributions. This assumption becomes more valid for asymptotically large number of cells and users.
\end{rem}
\section{Eigenvalue Distribution Analysis}
\label{sec: eigenvalue distribution analysis}

As it can be seen, the generic channel model of eq. \eqref{eq: generic channel model} and its ergodic capacity in eq. \eqref{eq: generic capacity} describe a wide range of multiantenna and multiple-access channels impaired by additive white Gaussian noise, flat fading, path loss and cochannel interference. In order to tackle this problem analytically, we resort to asymptotic analysis which entails that the dimensions of the channel matrices grow to infinity assuming proper normalizations. It has already been shown in many occasions that asymptotic analysis yields results which are also valid for finite dimensions \cite{Tulino04,Lozano2002,Martin2004}. In other words, the capacity expression converges quickly to a deterministic value as the number of channel matrix dimensions increases. Eq. \eqref{eq: generic capacity} can be written asymptotically as: 
\begin{align}
\mathrm{C}&=\lim_{K,M,N\rightarrow\infty}\frac{1}{K}\mathbb{E}\left[\log\det\left(\mathbf{I}+\mu\mathbf{HH}^H\left(\mathbf{I}+\nu\mathbf{H}_\mathrm{I}\mathbf{H}_\mathrm{I}^H\right)^{-1}\right)\right]\nonumber\\
&=\lim_{K,M,N\rightarrow\infty}\mathbb{E}\left[\frac{1}{K}\sum_{i=1}^K \log\left(1+\lambda_i\left(\mathbf{K}\right)\right)\right]\nonumber\\
&=\int_0^\infty \log\left(1+x\right)f^\infty_{\mathbf{K}}\left(x\right)\mathrm{d}x,
\label{eq: capacity aepdf}
\end{align}
where $\lambda_i\left(\mathbf{K}\right)$ is the $i$th eigenvalue of matrix $\mathbf{K}$ and $f^\infty_{\mathbf{K}}$ is the a.e.p.d.f. of $\mathbf{K}$.
It should be noted that while the channel dimensions $K,M,N$ grow to infinity the ratios of horizontal to vertical dimensions $\beta,\gamma$ are kept constant. More importantly, TSNR and TINR $\mu,\nu$ grow small as the transmit dimensions $M,N$ grow large in order to guarantee that the system TSNR and TINR $\tilde{\mu}=M\mu, \tilde{\nu}=N\nu$ remain constant and do not grow infinite in the context of asymptotic analysis. Based on the aforementioned conventions, the following auxiliary variables are defined:
\begin{align}
\tilde q&=\frac{\tilde{\mu} q}{\beta}={K\mu q}\nonumber\\
\tilde p&=\frac{\tilde{\nu} p}{\gamma}={K\nu p}\nonumber
\end{align}
To calculate the expression of eq. \eqref{eq: capacity aepdf}, it suffices to derive the a.e.p.d.f. of $\mathbf{K}$, which can be achieved through the principles of free probability theory \cite{Voiculescu83,Hiai2000,Hiai00,Bai1999} as described in the following paragraphs.
\begin{rem}
It should be noted that other techniques have been also used in recent literature for large random matrix analysis. Most notably, the replica analysis method \cite{Moustakas2007} and the deterministic equivalents method \cite{Hachem2007} have been applied in a range of wireless scenarios for deriving ergodic, outage capacities and precoding methods.  
\end{rem}

\subsection{Random Matrix Theory Preliminaries}
Let $f_\mathbf{X}(x)$ be the eigenvalue probability distribution function of a matrix \(\mathbf X.\)
\begin{df}
\label{def: n transform}
The $\eta$-transform of a positive semidefinite matrix \(\mathbf X\) is defined as\begin{equation}
\eta_{\mathbf X}\left(\gamma\right)=\int_0^\infty\frac{1}{1+\gamma x}f_\mathbf{X}(x)dx.
\end{equation}
\end{df}
\begin{df}
\label{def: Sigma transform}
The $\Sigma$-transform of a positive semidefinite matrix \(\mathbf X\) is defined as\begin{equation}
\Sigma_{\mathbf{X}}(x)=-\frac{x+1}{x}\eta^{-1}_{\mathbf{X}}(x+1).
\end{equation}
\end{df}
\begin{pro}
\label{pro: Sigma n}
The Stieltjes-transform of a positive semidefinite matrix \(\mathbf X\) can be derived by its $\eta$-transform using  \cite[Equation 2.48]{Tulino04}
\begin{equation}
\mathcal{S}_{\mathbf{X}}(x)=-\frac{\eta_{\mathbf{X}}(-1/x)}{x}.
\end{equation}
\end{pro}
\subsection{Free Probability Results}
The asymptotic capacity limit of this channel is calculated based on an asymptotic free probability approach which exploits the additive and multiplicative free convolution in the $R$- and $\Sigma$-transform domain respectively, as well as properties of the $\eta$ and Stieltjes transform. 
The derivation methodology in this paper can be summarized as follows:

\begin{enumerate}
        \item Derivation of a.e.p.d.f. of $\mathbf{N},\mathbf{\tilde N}$ through additive free convolution (Theorem \ref{thm: scaled marcenko pastur}) 
        \item Derivation of inverse $\eta$-transform of $\mathbf{M}$ through Cauchy integration (Theorem \ref{thm: eta transform M}) 
        \item Derivation of Stieltjes transform of $\mathbf{K}$ through multiplicative free convolution in the $S$-transform domain (Theorem \ref{thm: Stieltjes transform N}) 
        \item Calculation of $f^\infty_{\mathbf{K}}$ through Lemma \ref{pro: Stieltjes to aepdf}
        \item Integration based on eq. \eqref{eq: capacity aepdf} in order to calculate capacity
\end{enumerate}

\begin{thm}
\label{thm: scaled marcenko pastur}
The a.e.p.d.f. of $\mathbf{N},\mathbf{\tilde N}$ follows a scaled version of the Mar\v cenko-Pastur law, as long as $\mathbf{\Sigma},\mathbf{\Sigma}_\mathrm{I}$ are asymptotically row-regular.
\begin{proof}
Considering a Gaussian channel matrix \(\mathbf{G}\sim\mathcal{C}\mathcal{N}\left( \mathbf{0},\mathbf{I} \right)\), the empirical eigenvalue distribution  of \(\frac{1}{K}\mathbf{G}\mathbf{G}^H\)
converges almost surely (a.s.) to the non-random limiting eigenvalue distribution
of the Mar\v cenko-Pastur law \cite{Marcenko1967}, whose density function is given
by
\begin{align}
        f^\infty_{\frac{1}{K}\mathbf{G}\mathbf{G}^H}(x)&\stackrel{_{a.s.}}{\longrightarrow}\  f_{\mathrm{MP}}(x,\beta)
        \label{eq:marcenco pastur law}
\\f_{\mathrm{MP}}\left(x,\beta\right)&=\left(1-\beta\right)^+\delta\left(x\right)+\frac{\sqrt{\left(x-a\right)^+\left(b-x\right)^+}}{2\pi x}\nonumber
\end{align}
where $a=(1-\sqrt{\beta})^2,b=(1+\sqrt{\beta})^2$
and \(\eta\)-transform,  $S$-transform are given by \cite{Tulino04}
\begin{align}
\label{eq: n transform of Marcekno Pastur law}
\eta_{\mathrm{MP}}\left(x,\beta \right)&=1-\frac{\phi\left(x,\beta
\right)}{4 x}
\\\phi\left( x,\beta\right)&=\left(\sqrt{x\left( 1+\sqrt{\beta} \right)^2+1}-\sqrt{x\left( 1-\sqrt{\beta}\right)^2+1}\right)^{2}\nonumber
\\\Sigma_{\mathrm{MP}}\left(x,\beta \right)&=\frac{1}{\beta+x}
\end{align}
and \(\beta\) is the ratio of the horizontal to the vertical dimension of the \(\mathbf{G}\) matrix.

According to \cite{chatzinotas_letter,Chatzinotas_JWCOM,Chatzinotas}, the channel covariance matrix \(\mathbf{H}\mathbf{H}^H\) can be decomposed into a sum of unit rank matrices which are assumed asymptotic free\footnote{The reader is referred to \cite{chatzinotas_letter,Chatzinotas_JWCOM} for a complete proof of Theorem \ref{thm: scaled marcenko pastur}.}. Using additive free convolution in the $R$-transform domain, the empirical eigenvalue distribution  of \(\frac{1}{K}\mathbf{H}\mathbf{H}^H\)
is shown to converge almost surely (a.s.) to a scaled version
of the Mar\v cenko-Pastur law \cite{Marcenko1967}, as long as $\mathbf{\Sigma}$ is asymptotically row-regular
\begin{align}
        f^\infty_{\frac{1}{K}\mathbf{H}\mathbf{H}^H}(x)&\stackrel{_{a.s.}}{\longrightarrow}\  f_{\mathrm{MP}}(qx,\beta)
        \label{eq:scaled marcenco pastur law}
\end{align}
or equivalently
\begin{align}
\label{eq: scaled marcenco pastur law N}
        f^\infty_{\mathbf{N}}(x)&\stackrel{_{a.s.}}{\longrightarrow}\  f_{\mathrm{MP}}(\tilde{q}x,\beta)\\
        f^\infty_{\mathbf{\tilde N}}(x)&\stackrel{_{a.s.}}{\longrightarrow}\  f_{\mathrm{MP}}(\tilde{p}x,\gamma).
\end{align}
\end{proof}
\end{thm}

\begin{lem}
\label{lem: scaled marcenko pastur Sigma}
The $S$-transform of $\mathbf{N},\mathbf{\tilde N}$ converges almost surely (a.s.) to:
\begin{equation}
\Sigma_{\mathbf{N}}(x)\stackrel{_{a.s.}}{\longrightarrow}\  \frac{1}{\tilde q}\frac{1}{\beta+x}\\
\label{eq: sigma transform N}
\end{equation}
\begin{equation}
\Sigma_{\mathbf{\tilde N}}(x)\stackrel{_{a.s.}}{\longrightarrow}\  \frac{1}{\tilde p}\frac{1}{\gamma+x}
\label{eq: scaled marcenko pastur Sigma}
\end{equation}
\begin{proof}
According to Theorem \ref{thm: scaled marcenko pastur} and Definition \ref{def: n transform},
\begin{align}
\eta_{\mathbf{N}}(x)&\stackrel{_{a.s.}}{\longrightarrow}\  \eta_{\mathrm{MP}}(\tilde{q}x,\beta)\\
\eta^{-1}_{\mathbf{N}}(\tilde{q}x)&\stackrel{_{a.s.}}{\longrightarrow}\  \frac{1}{\tilde{q}}\eta_{\mathrm{MP}}^{-1}(x,\beta)
\end{align}
Using Definition \ref{def: Sigma transform} and eq. \eqref{eq: n transform of Marcekno Pastur law}, 
\begin{align}
\Sigma_{\mathbf{N}}(x)&=-\frac{x+1}{x}\eta^{-1}_{\mathbf{N}}(x+1)\nonumber\\
&\stackrel{_{a.s.}}{\longrightarrow}\  \frac{1}{\tilde{q}}\Sigma_{\mathrm{MP}}(x,\beta)\nonumber\\
&=\frac{1}{\tilde{q}}\frac{1}{\beta+x}.
\end{align}
A similar derivation can be followed for $\Sigma_{\mathbf{\tilde N}}(\nu)$.
\end{proof}
\end{lem}

\begin{thm}
\label{thm: eta transform M}
The inverse $\eta$-transform of $\mathbf{M}$ is given by:
\begin{equation}
\eta^{-1}_{\mathbf{M}}(x)=-{\frac { \left( x-1 \right)  \left( \sqrt {1+ \left( x-\gamma \right) ^{2}{\tilde{p}}^{2}+ \left( 2\,\gamma+2\,x \right) \tilde{p}}+1+ \left( \gamma-x \right) \tilde{p} \right) }{2x}}.
\label{eq: inverse eta transform M}
\end{equation}
\begin{proof}
See Appendix \ref{app: 1}.
\end{proof}
\end{thm}

\begin{thm}
\label{thm: Stieltjes transform N}
The inverse $\eta$-transform of $\mathbf{K}$ is given by:
\begin{equation}
\eta^{-1}_{\mathbf{K}}(x)=\frac{1}{\tilde q}\frac{1}{\beta+x-1}\eta^{-1}_{\mathbf{M}}(x)
\label{eq: inverse eta transform K}
\end{equation}
\begin{proof}
Assuming asymptotic freeness\footnote{Independent unitarily invariant matrices with compactly supported asymptotic spectra \cite[Example 2.45]{Tulino04}, such as Wishart and inverse Wishart matrices with identity covariance matrix \cite{Tempo2005}, are asymptotically free. The matrices $\mathbf N$ and $\mathbf M$ are independent, but the unitary invariance is not straightforward. More specifically, $\mathbf N$ is a Wishart matrix with a covariance matrix which depends on $\mathbf \Sigma$ and thus its distribution belongs to a more general class, called elliptically contoured matrix distribution \cite{Tempo2005}. In addition, \(\mathbf M\) is the inverse of a non-central  Wishart matrix as the sum of the channel covariance and the identity matrix. Nevertheless, the assumption of asymptotic freeness has been motivated by the accuracy of eigenvalue distributions and the fact that similar approximations
have been already investigated in an information theoretic
context, providing useful analytical insights and accurate numerical results \cite{Peacock2006,Hachem2002}.} between  matrices $\mathbf N$ and $\mathbf M$, the $\Sigma$-transform of $\mathbf{K}$ is given by multiplicative free convolution:
\begin{align}
\Sigma_{\mathbf{K}}(x)&=\Sigma_{\mathbf{N}}(x)\Sigma_{\mathbf{M}}(x)\nonumber\stackrel{_{(a)}}{\Longleftrightarrow}\\
\left(-\frac{x+1}{x}\right)\eta^{-1}_{\mathbf{K}}(x+1)&=\frac{1}{\tilde q}\frac{1}{\beta+x}\left(-\frac{x+1}{x}\right)\eta^{-1}_{\mathbf{M}}(x+1)\nonumber
\end{align}
where step $(a)$ combines Definition \ref{def: Sigma transform} and eq. \eqref{eq: sigma transform N}.
The variable substitution $y=x+1$ yields eq. \eqref{eq: inverse eta transform K}.
\end{proof}
\end{thm}

\begin{lem}
\label{pro: Stieltjes to aepdf}
The a.e.p.d.f. of $\mathbf{K}$ is obtained by determining the imaginary part of the Stieltjes transform \(\mathcal{S}\) for real arguments
\begin{equation}
f^{\infty}_\mathbf{K}(x)=\lim_{y\rightarrow0^+}\frac{1}{\pi}\mathfrak{I}\left\{ \mathcal{S}_\mathbf{K}(x+\mathrm{j}y)\ \right\},
\label{eq: limiting eigenvalue pdf}
\end{equation}
where the Stieltjes-transform \(\mathcal{S}\) of $\mathbf{K}$ is given by Property \ref{pro: Sigma n}
and the $\eta$-transform of $\mathbf{K}$ is calculated by inverting eq. \eqref{eq: inverse eta transform K}.
\end{lem}

\begin{lem}
\label{pro: Stieltjes to aepdf M}
The a.e.p.d.f. of $\mathbf{M}$ is obtained by determining the imaginary part of the Stieltjes transform \(\mathcal{S}\) for real arguments, where the Stieltjes-transform \(\mathcal{S}\) of $\mathbf{M}$ is given by Property \ref{pro: Sigma n} and the $\eta$-transform of $\mathbf{M}$ is given by eq. \eqref{eq: eta transform M} in Appendix \ref{app: 1}.
\end{lem}

\section{Numerical Results}
\label{sec: numerical results}
In order to verify the accuracy of the derived closed-form expressions and gain some insights on the capacity performance of the proposed generic model, a number of numerical results are presented in this section. In the following figures, solid lines are plotted based on closed-form expressions, while bars and circle points  are calculated based on Monte Carlo simulations. 

\subsection{A.e.p.d.f. Results}
\begin{figure}[htp]
  \begin{center}
    \subfigure[A.e.p.d.f. for  matrix $\mathbf{N}$]{\label{fig:1a}\includegraphics[width=0.5\textwidth]{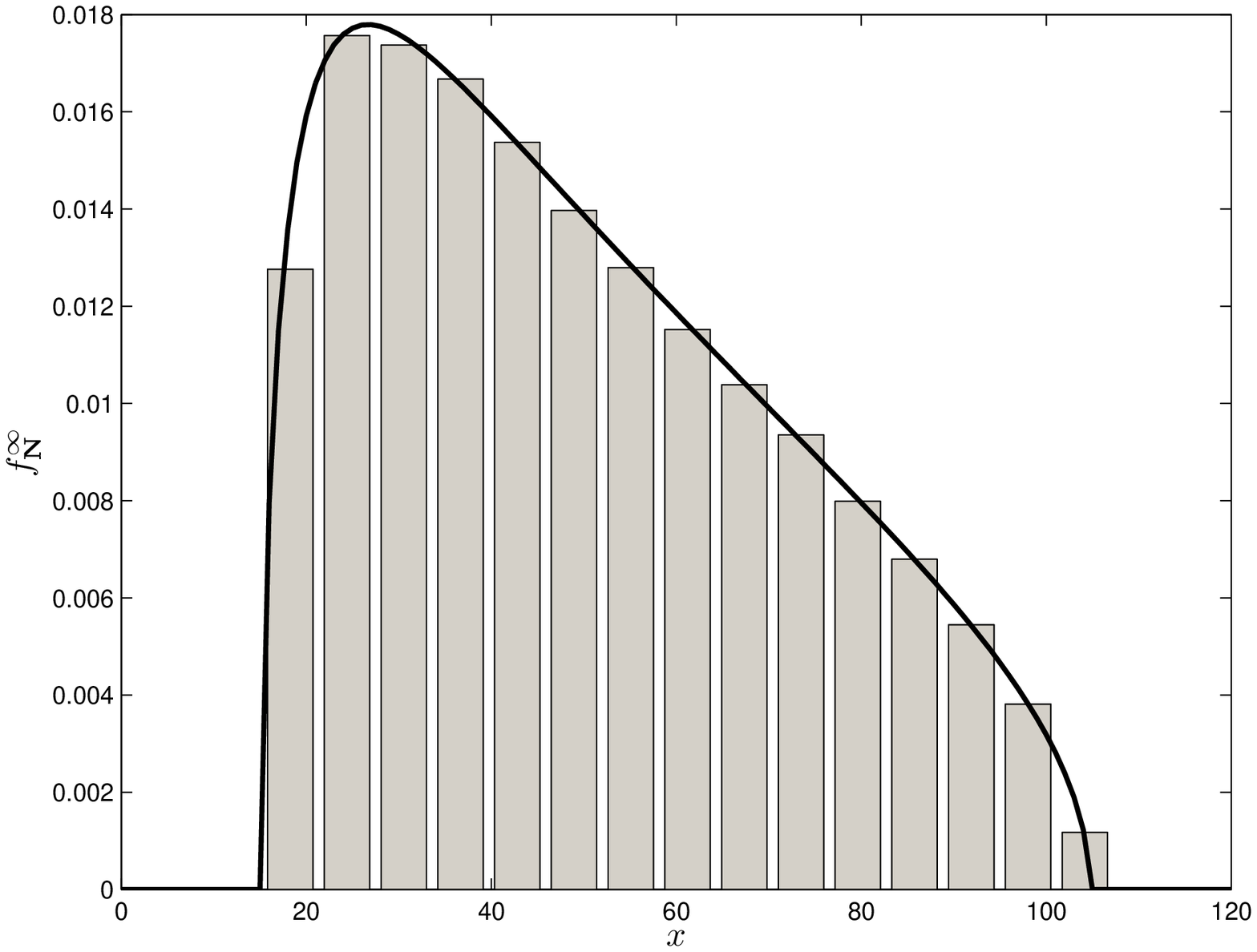}}\\
    \subfigure[A.e.p.d.f. for  matrix $\mathbf{M}$]{\label{fig:1b}\includegraphics[width=0.5\textwidth]{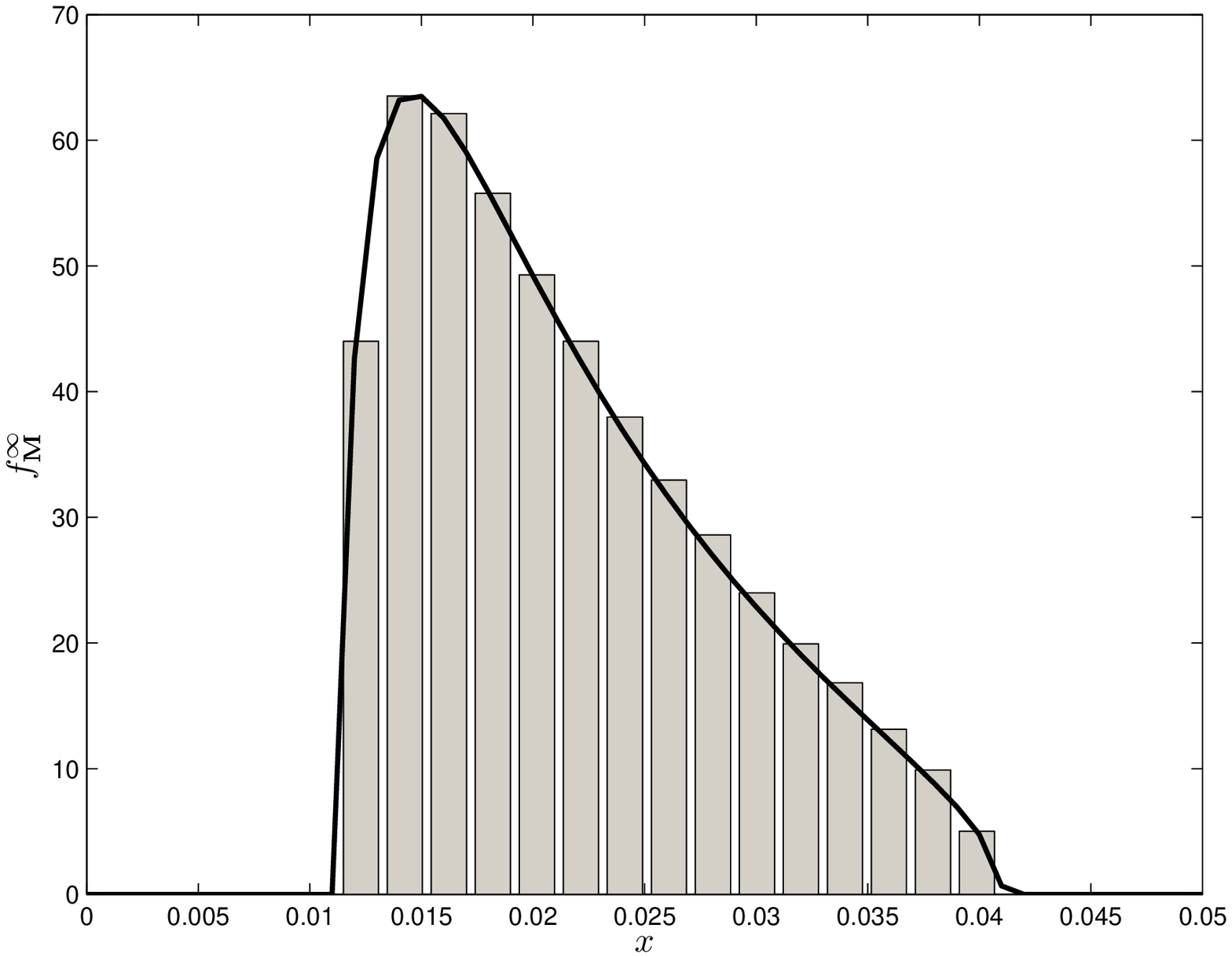}} \\
    \subfigure[A.e.p.d.f. for  matrix $\mathbf{K}$]{\label{fig:1c}\includegraphics[width=0.5\textwidth]{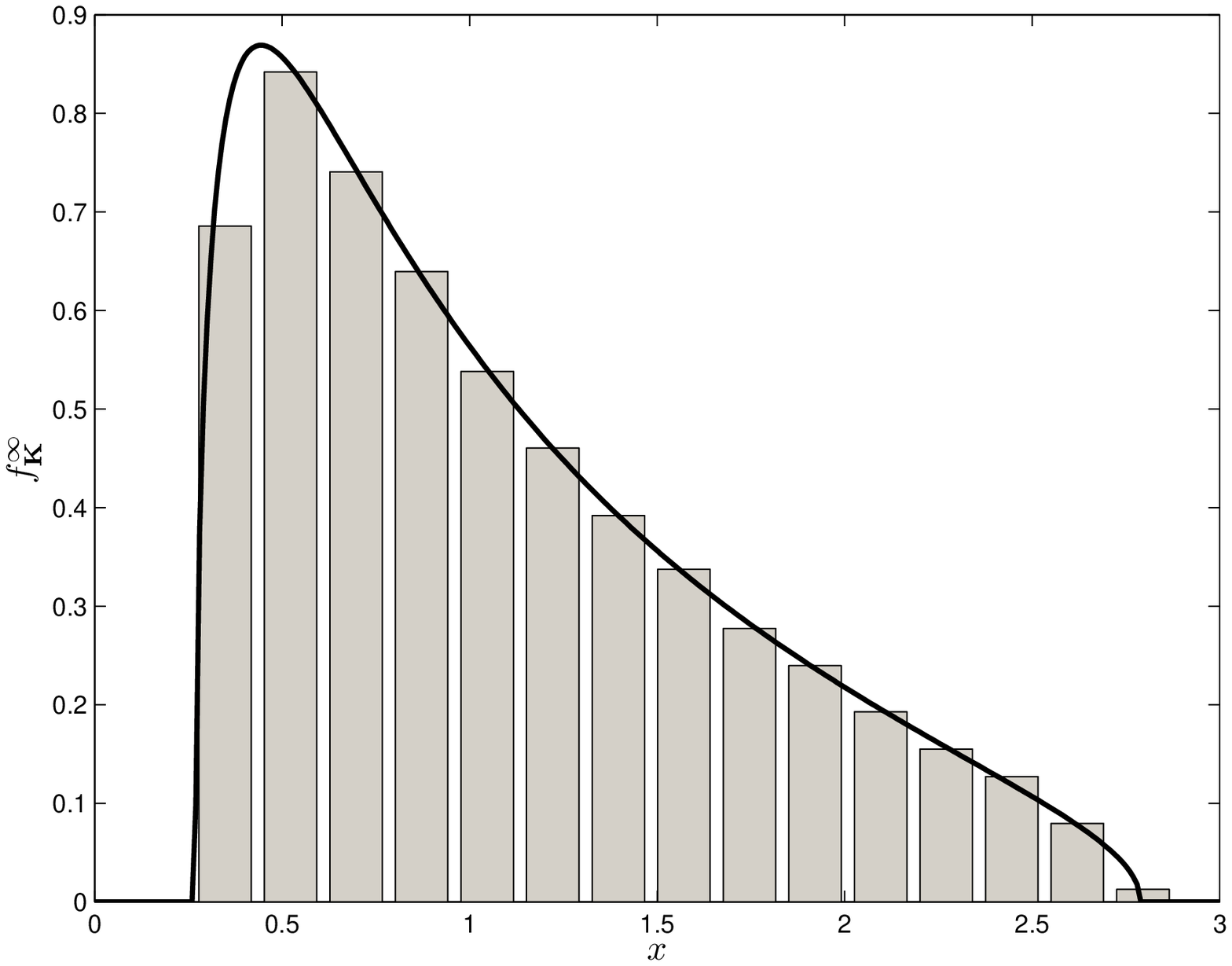}}
  \end{center}
  \caption{A.e.p.d.f. plots of matrices $\mathbf{K},\mathbf{M},\mathbf{N}$. Parameters: $\beta=5,\gamma=10,\nu q=10,\mu p=5$.}
  \label{fig: aepdf bar plots}
\end{figure}

The accuracy of the derived closed-form expressions for matrices $\mathbf{K},\mathbf{M},\mathbf{N}$ is depicted in Figure \ref{fig: aepdf bar plots}. The solid line in subfigure \ref{fig:1a} is drawn using Theorem \ref{thm: scaled marcenko pastur} and specifically eq. \eqref{eq: scaled marcenco pastur law N}, in subfigure \ref{fig:1b} using Lemma \ref{pro: Stieltjes to aepdf M} and in subfigure \ref{fig:1c} Lemma \ref{pro: Stieltjes to aepdf}. The histograms denote the p.d.f. of matrices $\mathbf{K},\mathbf{M},\mathbf{N}$ calculated numerically based on Monte Carlo simulations. More specifically, the matrices \(\mathbf G\) and \(\mathbf G_\mathrm I\) are generated using independent identically distributed (i.i.d.) complex circularly symmetric (c.c.s.) elements for \(10^3\) fading instances and the matrices \(\mathbf \Sigma\) and \(\mathbf \Sigma_\mathrm I\) using the following simplified model for diminishing off-diagonals:
\begin{equation}
\sigma_{i,j}=\frac{1}{\sqrt{1+\left\vert i-j\right\vert}},
\end{equation}
where $\sigma_{i,j}$ is the $(i,j)$th matrix element of \(\mathbf \Sigma\) or \(\mathbf \Sigma_\mathrm I\).
It can be seen that there is a perfect agreement between the two sets of results.

\begin{figure}[htp]
  \begin{center}
    \subfigure[Parameters: $\mu=\nu=\gamma=1$]{\label{fig:2a}\includegraphics[width=0.45\textwidth]{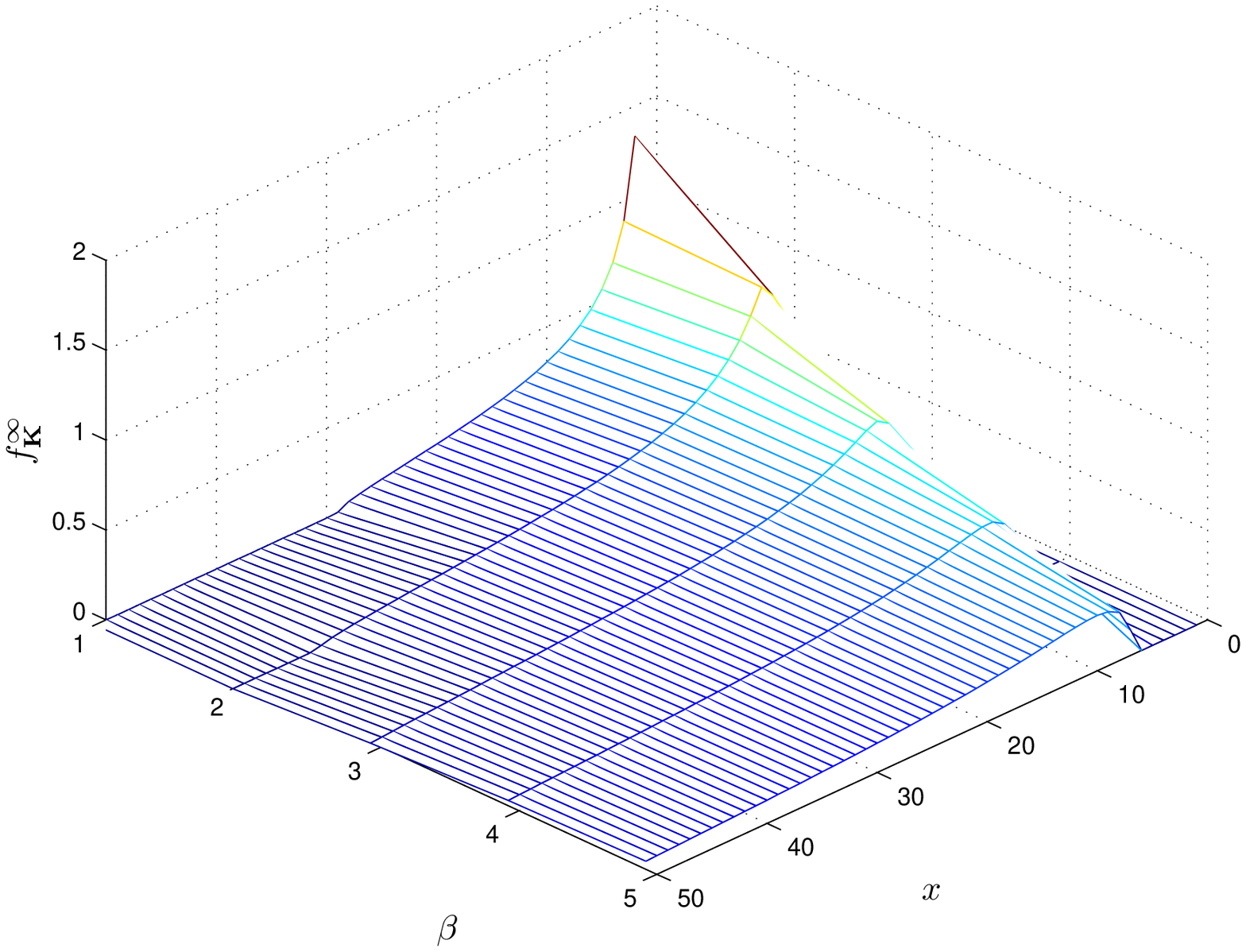}}
    \subfigure[Parameters: $\mu=\nu=\beta=1$]{\label{fig:2b}\includegraphics[width=0.45\textwidth]{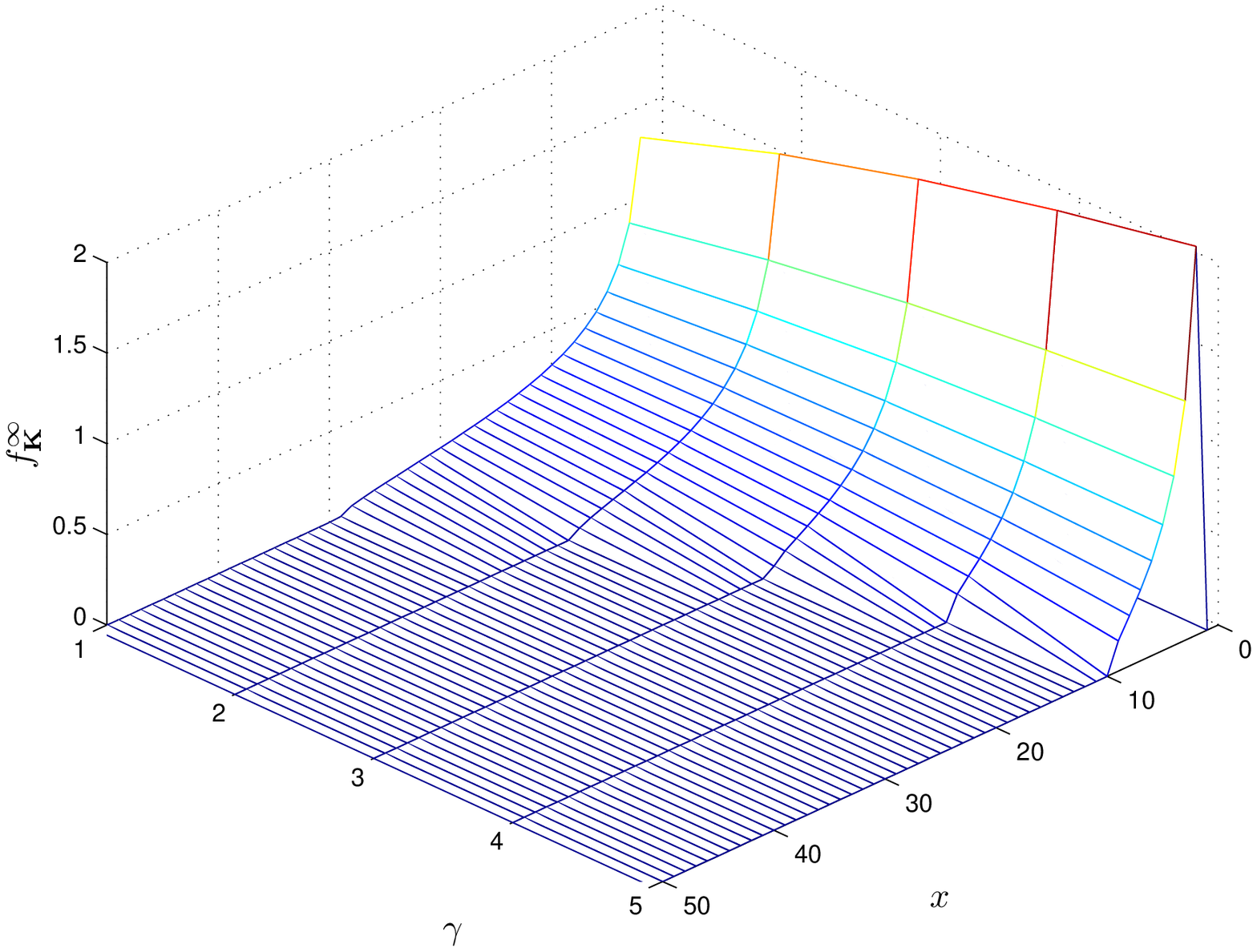}} \\
    \subfigure[Parameters: $\nu=\beta=\gamma=1$]{\label{fig:2c}\includegraphics[width=0.45\textwidth]{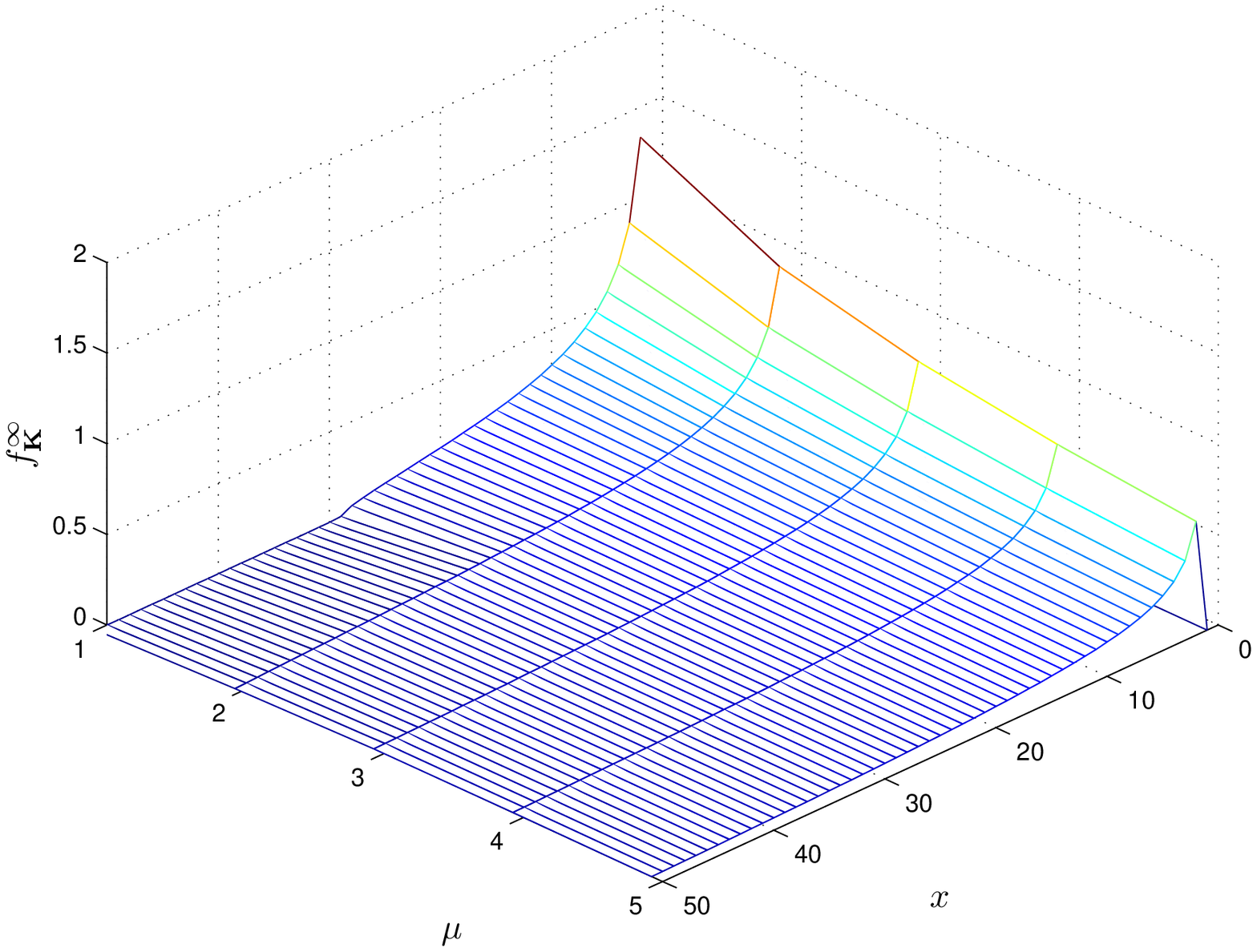}}
    \subfigure[Parameters: $\mu=\beta=\gamma=1$]{\label{fig:2d}\includegraphics[width=0.45\textwidth]{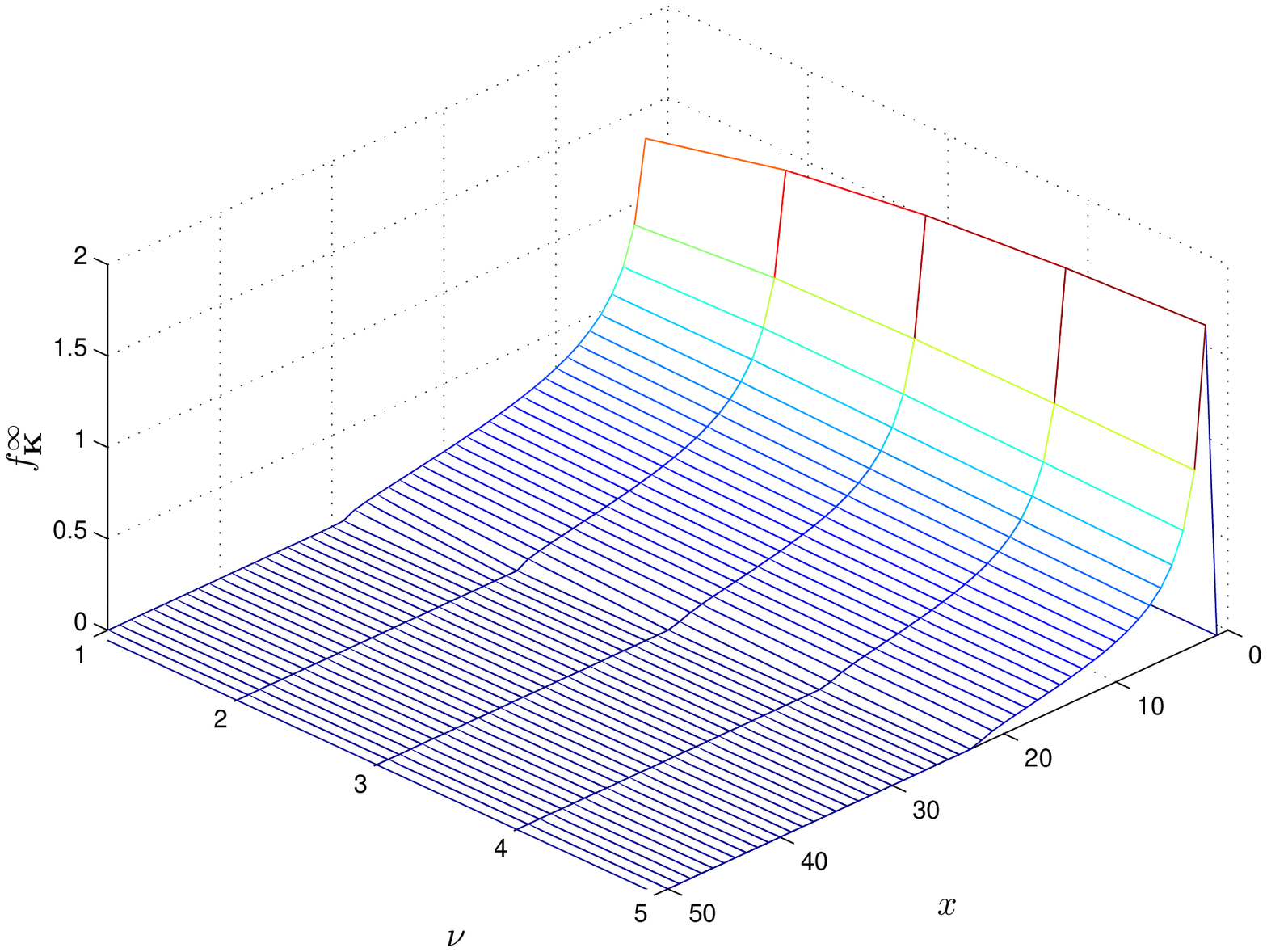}}
  \end{center}
  \caption{The effect of transmit power ($\mu,\nu$) and channel dimensions ($\beta,\gamma$) on the a.e.p.d.f. of matrix $\mathbf{K}$.}
  \label{fig: aepdf 3D plots}
\end{figure}

Figure \ref{fig: aepdf 3D plots} depicts the effect of transmit power ($\mu,\nu$) and channel dimensions ($\beta,\gamma$) on the a.e.p.d.f. of matrix $\mathbf{K}$. As it can be observed, the a.e.p.d.f. has four degrees of freedom, each one contributing to the shape of the curve. The final shape would be determined by combining those four contributions, namely the transmit power of desired dimensions $\mu$, the transmit power of interfering dimensions $\nu$, desired transmit over receive dimensions $\beta$ and interfering transmit over receive dimensions $\gamma$. The effect of the quantities $q,p$ is similar to that of $\mu,\nu$ and it is not depicted for the sake of conciseness. 

\subsection{Capacity Results}
\begin{figure}
        \centering
                \includegraphics[width=0.8\textwidth]{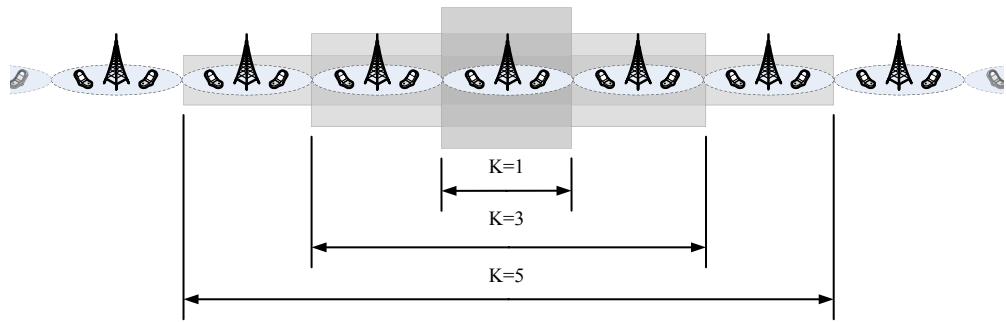}
        \caption{Graphical representation of a linear cellular array with cooperating BS clusters.}
        \label{fig: cellular array}
\end{figure}

\begin{figure}
        \centering
                \includegraphics[width=0.5\textwidth]{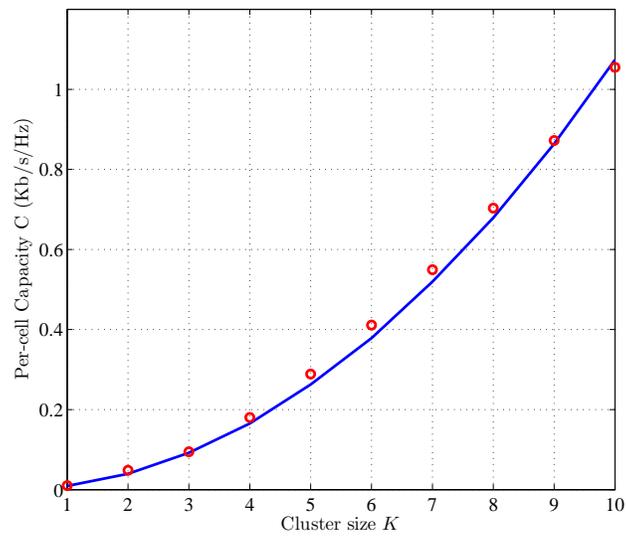}
        \caption{Per-cell capacity scaling vs. the cluster size $K$.}
        \label{fig: capacity scaling}
\end{figure}

In this paragraph, the focus is on cooperating BS clusters in the context of a linear cellular array with single-antenna BSs / UTs and wideband transmission scheme, as depicted in Figure \ref{fig: cellular array}. However, the presented closed-form expressions can be straightforwardly applied to planar arrays and multiple-antenna BSs and UTs. The parameters used for producing the capacity results are presented in Table \ref{tab:practicalparas}. The transmit power $P_T$, the number of UTs per cell $\beta$ and the cell size $R$ are kept constant while the number of cells $K$ participating in the cluster varies. The simulated system considers $50$ cells in total and $10^3$ Monte Carlo (MC) iterations. The UTs are assumed to be distributed on a regular grid across the system coverage span. TSNR and TINR are calculated as follows:
\begin{equation}
\mu=\nu=\frac{P_T}{N_0 B},
\end{equation}
while a power-law path loss model \cite{Letzepis,Ong2007,Chatzinotas_Chapter1} is employed for constructing the variance profile matrices $\mathbf{\Sigma},\mathbf{\Sigma}_\mathrm{I}$:
\begin{equation}
\sigma_{i,j}=\sqrt{P_0}\left(1+\frac{d_{i,j}}{d_0}\right)^{-\frac{n}{2}},
\end{equation}
where $\sigma_{i,j}$ is the $(i,j)$th matrix element and $d_{i,j}$ is the spatial distance between the $i$th receive dimension and the $j$th transmit dimension.
The quantities $q,p$ can be calculated either numerically or based on closed-form expressions as described in \cite{Chatzinotas_Chinacom,Chatzinotas_SPAWC} for linear and planar cellular arrays respectively.  It should also be noted that scaling the cluster size $K$ affects the structure of matrices $\mathbf{\Sigma},\mathbf{\Sigma}_\mathrm{I}$ and as a result the quantities $q,p$.
In this direction, the capacity plot versus cluster size in Figure \ref{fig: capacity scaling} includes two sets of results. The circle points denote values calculated based on simulating the channel matrices and averaging using eq. \eqref{eq: generic capacity}. The solid line is plotted using eq. \eqref{eq: capacity aepdf} and Theorem \ref{pro: Stieltjes to aepdf}. As it can be seen, the potential spectral efficiency of cooperating BS clusters is very high even for moderate cluster sizes.

\section{Conclusion}
\label{sec: conclusion}
In this paper, a generic multiantenna channel model was introduced, incorporating Gaussian noise, flat fading, path loss and cochannel interference. The proposed model generalizes and extends previous models in the literature by considering profile matrices which shape the variance of the fading coefficients for both desired and interfering signal dimensions. As a result, it is possible to describe a wide range of cochannel interference scenarios, including single-user and multi-user MIMO, as well as cooperating BS clusters, also known as distributed antenna systems. Based on a free probability approach, the asymptotic eigenvalue distribution has been derived, resulting in closed-form expressions which depend on TSNR, TINR, channel dimensions and norms of the variance profile matrices. Furthermore, the derivations herein have demonstrated how complicated channel matrix expressions can be tackled by additive and multiplicative free convolution in the $R$- and $S$-transform domain respectively, as well as properties of the $\eta$ and Stieltjes transform. Finally, the derived a.e.p.d.f. of the matrix products in the \(\log\det\) formula was utilized to calculate the capacity of cooperating BS clusters by varying the size of the cluster and to provide an estimation of the resulting spectral efficiencies.

\appendices
\section{Proof of Theorem \ref{thm: eta transform M}}
\label{app: 1}

Since the a.e.p.d.f. $f^\infty_\mathbf{\tilde N}(x)$ is known, the a.e.p.d.f. of $f^\infty_\mathbf{M}(y)$ can be calculated considering that  $y=(1+x)^{-1}$, where $y$ and $x$ represent the eigenvalues of $\mathbf{M}$ and $\mathbf{\tilde N}$ respectively:
\begin{align}
f^\infty_\mathbf{M}(y(x))&=\left| \frac{1}{y'(y^{-1}(x))} \right| \cdot f^\infty_\mathbf{\tilde N}(y^{-1}(x))\nonumber\\
&=x^{-2}f^\infty_\mathbf{\tilde N}\left(\frac{1-x}{x}\right).
\end{align}
Following the definition of $\eta$-transform \cite[Definition 2.11]{Tulino04},\cite{Bai09}:
\begin{align}
\eta_\mathbf{M}(\psi)&=\int_{-\infty}^{{+\infty}}\frac{1}{1+\psi x}f^\infty_\mathbf{M}(x)dx\nonumber\\
&=\int_{-\infty}^{{+\infty}}\frac{1}{1+\psi x}\frac{1}{x^2}{f}_\mathbf{\tilde N}\left(\frac{1-x}{x}\right)dx\nonumber\\
&\stackrel{_{(a)}}{=}-\int_{+\infty}^{{-\infty}}\frac{w+1}{1+\psi+w}{f}_\mathbf{\tilde N}(w)dw\nonumber\\
&=\int_{-\infty}^{{+\infty}}\frac{w+1}{1+\psi+w}{f}_\mathbf{\tilde N}(w)dw\nonumber\\
&=\int_{\tilde{a}}^{\tilde{b}}\frac{w+1}{1+\psi+w}\frac{1}{2\pi w\tilde{p}}\sqrt{(b-w)(x-w)}dw\nonumber\\
&\stackrel{_{(b)}}{=}-\gamma\int_0^\pi\frac{2}{\pi}\frac{1+\tilde{p}(1+\gamma+2\sqrt{\gamma}\cos \omega)}{(1+\gamma+2\sqrt{\gamma}\cos \omega)(1+\psi+\tilde{p}(1+\gamma+2\sqrt{\gamma}\cos \omega))}\sin^2\omega d\omega\nonumber\\
&=-\gamma\frac{1}{\pi}\int_0^{2\pi}\frac{(1+\tilde{p}(1+\gamma+\sqrt{\gamma}(e^{i\omega}+e^{-i\omega})))((e^{i\omega}-e^{-i\omega})/2i)^2}{(1+\gamma+\sqrt{\gamma}(e^{i\omega}+e^{-i\omega}))(1+\psi+\tilde{p}(1+\gamma+\sqrt{\gamma}(e^{i\omega}+e^{-i\omega})))}d\omega\nonumber\\
&\stackrel{_{(c)}}{=}\gamma\frac{1}{4i\pi}\oint_{\left|\zeta\right|=1}\frac{(1+\tilde{p}(1+\gamma+\sqrt{\gamma}(\zeta+\zeta^{-1})))(\zeta-\zeta^{-1})^2}{\zeta(1+\gamma+\sqrt{\gamma}(\zeta+\zeta^{-1}))(1+\psi+\tilde{p}(1+\gamma+\sqrt{\gamma}(\zeta+\zeta^{-1})))}d\zeta\nonumber\\
&=\gamma\frac{1}{4i\pi}\oint_{\left|\zeta\right|=1}\frac{((1+\tilde{p}(1+\gamma))\zeta+\sqrt{\gamma}\tilde{p}(\zeta^2+1))(\zeta^2-1)^2}{\zeta^2((1+\gamma)\zeta+\sqrt{\gamma}(\zeta^2+1))(\zeta(1+\psi+\tilde{p}(1+\gamma))+\sqrt{\gamma}\tilde{p}(\zeta^2+1))}d\zeta
\label{eq: eta derivation}
\end{align}
where $\tilde a=\tilde{p}(1-\sqrt{\gamma})^2$ and $\tilde b=\tilde{p}(1+\sqrt{\gamma})^2$. Step $(a)$ requires the variable substitution $x=1/(w+1)$, $dx=-1/(1+w)^{2}dw$, step $(b)$ requires $w=1+\gamma+2\sqrt{\gamma}\cos \omega$, $dw=2\sqrt{\gamma}(-\sin\omega)d\omega$ and step $(c)$ $\zeta=e^{i\omega}$,  $d\zeta=i\zeta d\omega$.

Subsequently, a Cauchy integration is performed by calculating the poles $\zeta_i$ and residues $\rho_i$ of eq. \eqref{eq: eta derivation}:
\begin{align}
\zeta_0,1&=0,\nonumber\\
\zeta_{2,3}&=\frac{-(1+\gamma)\pm(1-\gamma)}{2\sqrt{\gamma}},\nonumber\\
\zeta_{4,5}&={\frac {- \left( 1+\gamma \right) \tilde{p}-1-\psi\pm\sqrt { \left( \gamma-1 \right) ^{2}{\tilde{p}}^{2}+2\,\left( 1+\psi \right)  \left( 1+\gamma \right) \tilde{p}+\left( 1+\psi \right) ^{2}}}{2\tilde{p}\sqrt {\gamma}}},\nonumber\\
\rho_{0}&=-\frac{\tilde{p}+\tilde{p}\gamma+\psi}{\tilde{p}\gamma},\nonumber\\
\rho_{1}&=\frac{1}{\sqrt{\gamma}},\nonumber\\
\rho_{2,3}&=\pm\frac{\gamma-1}{\gamma+\psi \gamma},\nonumber
\end{align}
\begin{equation}
\textstyle{\rho_{4,5}=\pm
{\frac { \left(  -\left(  \left( 1+\gamma \right) \tilde{p}+1+\psi \right) \sqrt { \left( \gamma-1 \right) ^{2}{\tilde{p}}^{2}+2\, \left( 1+\psi \right)  \left( 1+\gamma \right) \tilde{p}+ \left( 1+\psi \right) ^{2}}+ \left( \gamma-1 \right) ^{2}{\tilde{p}}^{2}+2\, \left( 1+\psi \right)  \left( 1+\gamma \right) \tilde{p}+ \left( 1+\psi \right) ^{2} \right) \psi}{\tilde{p}\gamma\, \left( 1+\psi \right)  \left( -\sqrt { \left( \gamma-1 \right) ^{2}{\tilde{p}}^{2}+2\, \left( 1+\psi \right)  \left( 1+\gamma \right) \tilde{p}+ \left( 1+\psi \right) ^{2}}+ \left( 1+\gamma \right) \tilde{p}+1+\psi \right) }}\nonumber}
\label{eq:}
\end{equation}
Using the residues which are located within the unit disk, the Cauchy integration yields:
\begin{equation}
\eta_\mathbf{M}(\psi)=-\frac{\gamma}{2}(\rho_0+\rho_2+\rho_5)\nonumber
\end{equation}
\begin{align}
\textstyle{
=\frac { -\left( {\psi}^{2}+ \left( 1+ \left( 1+\gamma \right) \tilde{p} \right) \psi+\tilde{p} \right) \sqrt {\tilde{p}^{2} \left( -1+\gamma \right) ^{2}+2\, \left(1+\psi \right)  \left( \gamma+1 \right) \tilde{p}+ \left( 1+\psi \right) ^{2}}+{\psi}^{3}+ \left( 2+ \left( 2+2\,\gamma \right) \tilde{p} \right) {\psi}^{2}+ \left( 1+ \left( 1+{\gamma}^{2} \right) \tilde{p}^{2}+ \left( 2\,\gamma+3 \right) \tilde{p} \right) \psi+ \left( \gamma+1 \right) \tilde{p}^{2}+\tilde{p}}{ \left( 1+\psi \right)  \left( -\sqrt {\tilde{p}^{2} \left( -1+\gamma \right) ^{2}+2\, \left( 1+\psi \right)  \left( \gamma+1 \right) \tilde{p}+ \left( 1+\psi \right) ^{2}}+\psi+1+ \left( \gamma+1 \right) \tilde{p} \right)}
}
\label{eq: eta transform M}
\end{align}
Inversion yields eq. \eqref{eq: inverse eta transform M}.



\ifCLASSOPTIONcaptionsoff
  \newpage
\fi



\bibliographystyle{IEEEtran}
\bibliography{IEEEabrv,references,journals,books,conferences,thesis}

\begin{thebibliography}{10}
\providecommand{\url}[1]{#1}
\csname url@samestyle\endcsname
\providecommand{\newblock}{\relax}
\providecommand{\bibinfo}[2]{#2}
\providecommand{\BIBentrySTDinterwordspacing}{\spaceskip=0pt\relax}
\providecommand{\BIBentryALTinterwordstretchfactor}{4}
\providecommand{\BIBentryALTinterwordspacing}{\spaceskip=\fontdimen2\font plus
\BIBentryALTinterwordstretchfactor\fontdimen3\font minus
  \fontdimen4\font\relax}
\providecommand{\BIBforeignlanguage}[2]{{%
\expandafter\ifx\csname l@#1\endcsname\relax
\typeout{** WARNING: IEEEtran.bst: No hyphenation pattern has been}%
\typeout{** loaded for the language `#1'. Using the pattern for}%
\typeout{** the default language instead.}%
\else
\language=\csname l@#1\endcsname
\fi
#2}}
\providecommand{\BIBdecl}{\relax}
\BIBdecl

\bibitem{Blum2003}
R.~Blum, ``{MIMO} capacity with interference,'' \emph{{IEEE} J. Select. Areas
  Commun.}, vol.~21, no.~5, pp. 793--801, June 2003.

\bibitem{Lozano2002}
A.~Lozano and A.~Tulino, ``Capacity of multiple-transmit multiple-receive
  antenna architectures,'' \emph{Information Theory, IEEE Transactions on},
  vol.~48, no.~12, pp. 3117 -- 3128, dec 2002.

\bibitem{Moustakas2003}
A.~Moustakas, S.~Simon, and A.~Sengupta, ``{MIMO} capacity through correlated
  channels in the presence of correlated interferers and noise: a (not so)
  large {N} analysis,'' \emph{{IEEE} Trans. Inf. Theory}, vol.~49, no.~10, pp.
  2545--2561, Oct 2003.

\bibitem{Choi2008}
W.~Choi and J.~Andrews, ``The capacity gain from intercell scheduling in
  multi-antenna systems,'' \emph{{IEEE} Trans. Wireless Commun.}, vol.~7,
  no.~2, pp. 714--725, Feb 2008.

\bibitem{Somekh2007}
O.~Somekh, B.~Zaidel, and S.~Shamai, ``Sum rate characterization of joint
  multiple cell-site processing,'' \emph{{IEEE} Trans. Inf. Theory}, vol.~53,
  no.~12, pp. 4473--4497, Dec 2007.

\bibitem{Tulino04}
A.~M. Tulino and S.~{Verd\' u}, ``Random matrix theory and wireless
  communications,'' \emph{Commun. Inf. Theory}, vol.~1, no.~1, pp. 1--182,
  2004.

\bibitem{Chatzinotas_arxiv}
\BIBentryALTinterwordspacing
S.~Chatzinotas and B.~Ottersten, ``Interference alignment for clustered
  multicell joint decoding,'' \emph{{IEEE} Trans. Wireless Commun.}, 2010,
  submitted. [Online]. Available: \url{http://arxiv.org/abs/1007.4112}
\BIBentrySTDinterwordspacing

\bibitem{Foschini1998}
G.~J. Foschini and M.~J. Gans, ``On limits of wireless communications in a
  fading environment when using multiple antennas,'' \emph{Wirel. Pers.
  Commun.}, vol.~6, no.~3, pp. 311--335, 1998.

\bibitem{Wyner1994}
A.~Wyner, ``Shannon-theoretic approach to a {Gaussian} cellular multiple-access
  channel,'' \emph{{IEEE} Trans. Inf. Theory}, vol.~40, no.~6, pp. 1713--1727,
  Nov 1994.

\bibitem{Somekh2000}
O.~Somekh and S.~Shamai, ``Shannon-theoretic approach to a {Gaussian} cellular
  multiple-access channel with fading,'' \emph{{IEEE} Trans. Inf. Theory},
  vol.~46, no.~4, pp. 1401--1425, Jul 2000.

\bibitem{Dai2004}
H.~Dai, A.~Molisch, and H.~Poor, ``Downlink capacity of interference-limited
  {MIMO} systems with joint detection,'' \emph{{IEEE} Trans. Wireless Commun.},
  vol.~3, no.~2, pp. 442--453, March 2004.

\bibitem{Chatzinotas_Chapter1}
S.~Chatzinotas, M.~Imran, and C.~Tzaras, \emph{Cooperative Wireless
  Communications}.\hskip 1em plus 0.5em minus 0.4em\relax Auerbach
  Publications, Taylor \& Francis Group, 2009, ch. Capacity Limits in
  Cooperative Cellular Systems.

\bibitem{Martin2004}
C.~Martin and B.~Ottersten, ``Asymptotic eigenvalue distributions and capacity
  for {MIMO} channels under correlated fading,'' \emph{{IEEE} Trans. Wireless
  Commun.}, vol.~3, no.~4, pp. 1350--1359, Jul 2004.

\bibitem{Voiculescu83}
D.~Voiculescu, ``Asymptotically commuting finite rank unitary operators without
  commuting approximants,'' \emph{Acta Sci. Math.}, vol.~45, pp. 429--431,
  1983.

\bibitem{Hiai2000}
F.~Hiai and D.~Petz, ``Asymptotic freeness almost everywhere for random
  matrices,'' \emph{Acta Sci. Math. (Szeged)}, vol.~66, pp. 801--826, 2000.

\bibitem{Hiai00}
------, ``The semicircle law, free random variables and entropy,''
  \emph{Mathematical Surveys and Monographs}, vol.~77, 2000.

\bibitem{Bai1999}
Z.~D. Bai, ``Methodologies in spectral analysis of large dimensional random
  matrices, a review,'' \emph{Statistica Sinica}, vol.~9, pp. 611--677, 1999.

\bibitem{Moustakas2007}
A.~Moustakas and S.~Simon, ``On the outage capacity of correlated multiple-path
  {MIMO} channels,'' \emph{{IEEE} Trans. Inf. Theory}, vol.~53, no.~11, pp.
  3887--3903, Nov. 2007.

\bibitem{Hachem2007}
\BIBentryALTinterwordspacing
W.~Hachem, P.~Loubaton, and J.~Najim, ``Deterministic equivalents for certain
  functionals of large random matrices,'' \emph{Annals of Applied Probability},
  vol.~17, pp. 875--930, 2007. [Online]. Available:
  \url{http://arxiv.org/abs/math.PR/0507172}
\BIBentrySTDinterwordspacing

\bibitem{Marcenko1967}
V.~{Mar\v cenko} and L.~Pastur, ``Distributions of eigenvalues of some sets of
  random matrices,'' \emph{Math. USSR-Sb.}, vol.~1, pp. 507--536, 1967.

\bibitem{chatzinotas_letter}
S.~Chatzinotas, M.~Imran, and C.~Tzaras, ``On the capacity of variable density
  cellular systems under multicell decoding,'' \emph{{IEEE} Commun. Lett.},
  vol.~12, no.~7, pp. 496 -- 498, Jul 2008.

\bibitem{Chatzinotas_JWCOM}
S.~Chatzinotas, M.~Imran, and R.~Hoshyar, ``On the multicell processing
  capacity of the cellular {MIMO} uplink channel in correlated {Rayleigh}
  fading environment,'' \emph{{IEEE} Trans. Wireless Commun.}, vol.~8, no.~7,
  pp. 3704--3715, July 2009.

\bibitem{Chatzinotas}
S.~Chatzinotas, ``Information-theoretic capacity limits in multi-cell joint
  processing systems,'' Ph.D. dissertation, Center for Communication Systems
  Research, University of Surrey, Mar 2009, {A}vailable online at
  \url{http://wwwen.uni.lu/content/download/29816/358216/file/Chatzinotas_PhD_%
thesis.pdf}.

\bibitem{Tempo2005}
R.~Tempo, G.~Calafiore, and F.~Dabbene, \emph{Randomized Algorithms for
  Analysis and Control of Uncertain Systems}, ser. Communications and Control
  Engineering Series.\hskip 1em plus 0.5em minus 0.4em\relax London: Springer,
  2005.

\bibitem{Peacock2006}
M.~Peacock, I.~Collings, and M.~Honig, ``Asymptotic spectral efficiency of
  multiuser multisignature {CDMA} in frequency-selective channels,''
  \emph{{IEEE} Trans. Inf. Theory}, vol.~52, no.~3, pp. 1113--1129, Mar 2006.

\bibitem{Hachem2002}
W.~Hachem, ``Low complexity polynomial receivers for downlink {CDMA},'' in
  \emph{36th Asilomar Conference on Signals, Systems and Computers (ACSSC'02)},
  vol.~2, Nov 2002, pp. 1919--1923.

\bibitem{Letzepis}
N.~Letzepis, ``Gaussian cellular muptiple access channels,'' Ph.D.
  dissertation, Institute for Telecommunications Research, University of South
  Australia, Dec 2005.

\bibitem{Ong2007}
L.~Ong and M.~Motani, ``On the capacity of the single source multiple relay
  single destination mesh network,'' \emph{Ad Hoc Netw.}, vol.~5, no.~6, pp.
  786--800, 2007.

\bibitem{Chatzinotas_Chinacom}
S.~Chatzinotas, M.~Imran, and C.~Tzaras, ``The effect of user distribution on a
  cellular multiple-access channel,'' in \emph{Third International Conference
  on Communications and Networking in China (ChinaCom'08)}, Hangzhou, China,
  August 2008, pp. 95--99.

\bibitem{Chatzinotas_SPAWC}
S.~Chatzinotas, M.~A. Imran, and C.~Tzaras, ``Optimal information theoretic
  capacity of the planar cellular uplink channel,'' in \emph{IEEE 9th Workshop
  on Signal Processing Advances in Wireless Communications (SPAWC'08)},
  Pernambuco, Brazil, Jul 2008, pp. 196--200.

\bibitem{Bai09}
Z.~Bai and J.~Silverstein, \emph{Spectral Analysis of Large Dimensional Random
  Matrices}, 2nd~ed., ser. Springer Series in Statistics.\hskip 1em plus 0.5em
  minus 0.4em\relax New York: Springer, 2009.

\end{thebibliography}

\newpage

\begin{table}
\caption{Parameters for capacity results}
\centering
\begin{tabular}{c|c|l}
Parameter & Symbol & Value/Range (units)\\\hline
Cell Radius & \(R\) & $1\ Km$\ \\
Reference Distance  & \(d_{0}\) & $1\ m$ \\
Reference Path Loss  & \(P_{0}\) & $34.5\ dB$ \\
Path Loss Exponent  & \(n\) & $3.5\ $ \\
UTs per Cell & \(\beta\) & $10$ \\
Cluster Size& \(K\) & $1-10$ \\
Total number of cells& \(K+\frac{N}{\beta}\) & $50$ \\
UT Transmit Power & \(P_{T}\) & $200\  mW$ \\
Thermal Noise Density & \(N_0\) & $-169\  dBm/Hz$ \\
Channel Bandwidth & \(B\) & $5\  MHz$ \\
Number of MC iterations &   & $10^3$ \\\hline
\end{tabular}
\label{tab:practicalparas}
\end{table}

\end{document}